\documentclass[preprint,showpacs,preprintnumbers,amsmath,amssymb,epsfig]{revtex4}

\usepackage{graphicx}
\usepackage{dcolumn}
\usepackage{bm}

\begin{document}

\title{Compact magneto-optical sources of slow atoms}

\author{Yuri B. Ovchinnikov}
\affiliation{National Physical Laboratory, Hampton Road, Teddington, Middlesex, TW11 0LW, UK}

\email{yuri.ovchinnikov@npl.co.uk}

\date{\today}

\begin{abstract}
Three different configurations of compact magneto-optical sources of slow $^{87}$Rb atoms
(LVIS, 2D$^{+}$-MOT and 2D-MOT) were compared with each other at fixed geometry
of cooling laser beams. A precise control of the intensity balances between the four separate
transverse cooling laser beams provided a total continuous flux of cold atoms from the LVIS
and 2D$^{+}$-MOT sources about $8 \times 10^{9}$ \thinspace atoms/s at total laser power of
60 \thinspace mW.
The flux was measured directly from the loading rate of a 3D-MOT, placed 34 cm downstream from the sources.
Average velocities of the cooled atomic beam for the LVIS and 2D$^{+}$-MOT sources were
about 8.5 \thinspace m/s and 11 m/s respectively.
An essential advantage of the compact magneto-optical sources is that their background flux of
thermal atoms is two to three orders of the magnitude smaller than the flux of slow atoms.
\end{abstract}

\maketitle

\section{Introduction}
Efficient loading of atomic traps in an ultra-high-vacuum environment, like
UHV Magneto-Optical Traps (MOT) \cite{Raab} or magnetic guides \cite{Roos,Conroy},
requires intense beams of slow atoms.
The most intense source of slow atoms is a Zeeman slower \cite{Phillips} and
for Rb can provide fluxes of up to $10^{11}$ at/s at
velocities of 30 m/s. The main disadvantages of Zeeman slowers are their complexity, large size,
essential stray magnetic fields, large final divergence of cold atoms and high background
of thermal atoms. The general concept of atomic beam deceleration requires an initial thermal
beam with a flux, which exceeds the final flux of cold atoms by at least one order of magnitude.
Therefore many UHV experiments, which are sensitive to intense background flux of thermal atoms,
demand an additional in-vacuum mechanical shutter to block the thermal atomic beam. Introducing such a shutter
inside an UHV chamber, even in the high-pressure part, is a challenging problem, taking
into account that many of these experiments, like BEC experiments \cite{Cornell}, are sensitive to mechanical vibrations.

The concept of magneto-optical sources of atoms is quite different - to extract slow atoms from an
ultra-cold sample of atoms prepared in a magneto-optical trap. The advantages of this approach are very small
velocity and divergence of extracted atoms and low background of thermal atoms.

The main motivation of this work was to create a compact, robust, economic and efficient
magneto-optical source of cold atoms for loading an UHV MOT of a Rb fountain
standard of frequency.
The main demand on the source in that experiment was to have a high flux of cold atoms at
minimum background of thermal atoms. This goal was achieved by minimizing the length of the
source chamber and introducing an additional differential tube between the MOT-source and
UHV-MOT chambers.
An essential difference of our MOT-source, compared to previous ones \cite{Lu,Dieckmann,Conroy}, is that
for transverse cooling of atoms, four independent laser beams are used instead of retroreflected beams.
This allows us to get a perfect balance between these beams and achieve maximum flux of cold atoms
at smaller power of light.

In this paper, Section 2 describes the model of the MOT source, which is exploited to estimate the flux and
velocity of the beam of cold atoms produced by the source. In Section 3 the measurements of the main parameters
of three different configurations of the MOT source (LVIS, 2D$^{+}$-MOT and 2D-MOT) are
presented and compared to the theoretical results. In the concluding Section 4 the main results of the work are summarized.

\section{Model of the MOT-source}

Any continuous MOT-source of slow atoms consists of a magneto-optical trap with a dark channel in it, through which the
slow atoms are extracted. The first and the most straightforward version of such a low-velocity intense source (LVIS)
\cite{Lu} is based on a standard 3D-MOT, in which one of the six beams has a dark channel at its centre.
A 2D$^+$-MOT \cite{Dieckmann} is very similar to an LVIS except that it uses a two-dimensional quadruple magnetic field.
A pure 2D-MOT \cite{Schoser} uses only
four laser beams for transverse cooling of atoms and a two-dimensional transverse quadruple magnetic
field. As it was shown in \cite{Schoser}, the pure 2D-MOT source of short length is expected to be much less efficient than the compact
LVIS and 2D$^+$-MOT sources.

To estimate basic parameters of the sources a very simplified model of the MOT-source, based on a two-level model
of an atom, which excludes all stochastic processes related to spontaneous scattering of photons,
was applied.
The corresponding dissipative force, which describes the motion of an atom in the trap, is given
by

\begin{subequations}
\begin{eqnarray}
 \vec{F_{tot}}(x,y,z)=\vec{F_{x +}}+\vec{F_{x -}}
+\vec{F_{y +}}+\vec{F_{y -}}
+C(z)(\vec{F_{z +}}+A(x,y) \vec{F_{z -}}),\label{appa}\\
  \vec{F_{l \pm}}=\pm \frac{\hbar \vec{k} \Gamma}{2} \frac{s_{m,n}}{1+6 s_{m,n}+4(\delta_{l
  \pm}/\Gamma)^2},\label{appb}
\end{eqnarray}
\end{subequations}

where $l=\{x,y,z\}$ are three Cartesian coordinates of the MOT, $m$ and $n$ are two coordinates
complementary to $l$ of the same coordinate system,
$s_{m,n}=s_{0} e^{-2(m^{2}+n^{2})/w^{2}}$ is the saturation parameter of the atomic transition in the six cooling Gaussian
laser beams with waist radius $w$,
$\delta_{l \pm}=\Delta \mp \vec{k} \vec{v} \pm \mu' \nabla B_{l} l/\hbar$ is the
effective frequency detuning of the MOT's laser fields, $\Delta=\omega-\omega_{0}$ is the frequency detuning
of the laser field of frequency $\omega$ from the unperturbed atomic transition $\omega_{0}$,
$\nabla B_{l}$ is the gradient of magnetic field along the corresponding $l$-axes and
$\Gamma$ is a natural FWHM width of the atomic transition.

The orientation of the anti-Helmholtz coils along the y-axis
assumes that $\nabla B_{y}=-2 \nabla B_{x}=-2 \nabla B_{z}$. The
factor $\mu'= 5\mu_{B}/6$ is an averaged magnetic moment of the
$J=1/2 \rightarrow J=3/2$ atomic transition, where $\mu_{B}$ is
the Bohr magneton constant. The factor of six in front of the
saturation parameter in the denominator of the force is introduced
to take into account the cross-saturation of six laser beams.
Equation (1) is empirical but provides the right magnitude of the
force at low and high values of the saturation parameter. It is
also in reasonable agreement with our experimental observations.
The coefficient $A(x,y)=1-\exp[-2(x^{8}+y^{8})/a^{8}]$ in front of
the $\vec{F}_{z -}$ modelled the extraction channel of radius $a$
in one of the laser beams of the LVIS or the 2D$^+$-MOT. The
second factor $C(z)=1/(1+\exp[8(z-b)])$ modelled the
retroreflecting mirror of the source, which was setting up a
border for the intensity of the longitudinal laser beams at a
distance $z=b=2\,$cm from the centre of the trap. Both functions
$A(x,y)$ and $C(z)$ were chosen to be steep enough to provide
adequate results, but not too steep thus allowing the numerical
integration of the Newtonian equation of atomic motion in the
field of the force $\vec{F}_{tot}(x,y,z)$. Based on the model of
the trapping force presented, a numerical solution of a system of
differential equations, characterizing a classical motion of atoms
in the trap, was solved. Fig.$\,$1 illustrates a two-dimensional
motion of ten Rb atoms in the XOZ plane of the LVIS. All atoms
start at a distance $w$ from the centre of the trap with their
velocities within the capture range of the MOT. Here we used the
same parameters as in the experiment: $\Gamma=2\pi \times
6.0\,$MHz, $\nabla B_{y}=8\,$G/cm, $w=1.25\,$cm, $a=0.1\,$cm,
$b=2\,$cm, $s_{0}=3$, $\Delta=-3 \Gamma$. The two upper rectangles
simulate the retroreflecting mirror with a 2$\,$mm hole in its
centre.

This model makes it possible to estimate the maximum flux of the MOT-source and the velocity distribution of extracted atoms.
As it was shown in reference \cite{Lu}, when the extraction of atoms from the source is much faster than the collision rate
between them, the total flux of cold atoms $\Phi$ of the LVIS is simply equal to the loading rate of the trap R. Later on in the
paper \cite{Dieckmann} an additional factor, which is taking into account the losses due to collisions of cold atoms with
background thermal atoms, was introduced.
In our case the loading rate of the MOT-source was calculated as

\begin{equation}
\Phi=K_{coll} n_{87} S \iiint_{\vec{v}<\vec{v_{c}}} v_{\bot} g(\vec{v},T)\,d\vec{v},
\end{equation}

where $n_{87}\simeq 2.16 \times 10^{9}\,$at/cm$^{3}$ is the total
density of $^{87}Rb$ atoms, S is the area of the surface
surrounding the trap, $v_{\bot}$ is the component of the atomic
velocity, which is normal to the surface S and directed inwards
the trap and $g(\vec{v},T)$ is the Maxwell-Boltzmann probability
weighting factor for the atoms of velocity $\vec{v}$. The
integration was performed only over the velocity class of atoms
trapped in the MOT by the damping force introduced above. For
values of the saturation parameter $s_{0}$ between 0.5 and 8, the
corresponding maximum radial capture velocities of the MOT-source
range between 23 \thinspace m/s and 34 \thinspace m/s. The factor
$K_{coll}$ in equation (2) represents the reduction of the flux
due to collisions of cold rubidium atoms with thermal background
rubidium atoms. As in paper \cite{Dieckmann} we suppose that these
collisions mostly affect the cold atoms during their motion
towards the output of the MOT source. Therefore the loss factor
can be defined as $K_{coll}=\exp(-\Gamma_{beam} t_{out})$, where
$\Gamma_{beam}=\sigma_{eff} n_{Rb} \bar{v} \approx 415\,$s$^{-1}$
is the rate of the collisions between the cold atoms of the beam
with thermal Rb atoms, $t_{out} \approx 2\,$ms is the mean
travelling time of the captured atoms from the centre to the
output of the MOT. To calculate $\Gamma_{beam}$ we used the
effective collision cross section $\sigma_{eff}=2 \times
10^{-12}\,$cm$^2$ \cite{Dieckmann, Schoser}, the total density
$n_{Rb}$ of both the main isotopes of $Rb$ and average thermal
velocity of atoms $\bar{v}=270\,$m/s. We estimate the total
density as $n_{Rb}=n_{85}+n_{87}=7.7 \times 10^9\,$at/cm$^3$ from
the pressure of Rb vapor P$=3.17 \times 10^{-7}\,$mbar at our room
temperature T=296$\,$ K according to data of \cite{Nesmeyanov}.

The Fig.$\,$2 shows the calculated maximum flux of the LVIS source
as a function of the maximum saturation parameter $s_{0}$ of
cooling laser beams at fixed frequency detuning of $\Delta=-3
\Gamma$.

The calculated velocity distributions of atoms extracted from the LVIS and the 2D$^{+}$-MOT are shown in Fig.3.
To model the 2D$^{+}$-MOT the longitudinal gradient of the magnetic field $\nabla B_{z}$
was set to zero. One can see that the total fluxes of LVIS and 2D$^{+}$-MOT are nearly equal
to one another. On the other hand, the average velocity of slow atoms in the 2D$^{+}$-MOT
($\simeq 9\,$m/s) is slightly higher than the velocity of the LVIS beam ($\simeq 8\,$m/s).
The velocity of the LVIS is lower because of the Zeeman shift, which detunes the frequency of the
pushing beam $\vec{F}_{z+}$ out of resonance with atomic transition as atoms are moving towards the output of the source.
The model shows that the average velocity of extracted atoms strongly depends
on the distance between the center of the trap and its output border, formed by the retroreflecting mirror.
Obviously the longer acceleration distance leads to higher velocities of extracted atoms.

The model can also estimate the dependence of the
atomic flux on the power imbalance between the transverse
counterpropagating cooling laser beams of the MOT-source.
Fig.$\,$4 shows the calculated flux of the LVIS as a function of
the power imbalance
$(P_{x-}-P_{x+})/P_{x-}+P_{x+})=(P_{y-}-P_{y+})/P_{y-}+P_{y+})$ in
two pairs of the transverse laser beams. In these calculations a
maximum saturation parameter of $s_{0}=3$ and a frequency
detuning of $\Delta=-3 \Gamma$ were used. The atomic flux
decreases rapidly after the power imbalance in each of the two
pairs of transverse laser beams becomes larger than 4$\%$. The
trajectories of atoms in the LVIS, calculated for a power
imbalance of 5$\%$, are presented in Fig.$\,$5. One can see that
the imbalance of transverse light forces leads to deflection of
the extracted atoms from the dark channel and their re-trapping in
the MOT. Very similar behavior is observed also for the motion of
atoms in the 2D$^{+}$-MOT. Although in the 2D$^{+}$-MOT the
deflected atoms are not re-trapped, the divergence of the beam of
cold atoms increases dramatically. This model predicts also
that the sensitivity of the MOT-sources to the power imbalances
increases with decreasing diameter of the extraction
channel.

An estimate of the flux of thermal atoms for the retroreflecting mirror of 6$\,$mm thickness with a
2$\,$mm central hole gives the total background flux of both Rb isotopes, of about $1.9 \times 10^7\,$at/s.
This flux is nearly 3 orders of the magnitude smaller than the calculated flux of cold atoms at $s_{0}>3$.
The most critical parameter for reducing the background thermal flux is the length of the differential tube,
formed by the hole in the mirror.

\section{Experiment}

An experimental setup of a MOT-source is shown in Fig.\thinspace
6. A standard DN40 stainless steel cross with five standard
optical viewports was used as the body of the source. Inside the
output port of the cube an ensemble of retro-reflecting optics,
consisting of a gold-coated mirror with a $\lambda/4$-retardation
plate on top of it, was installed. Both elements were of 25$\,$mm
in diameter and had a 2$\,$mm hole in a centre. The mirror had a
width of 6$\,$mm and was attached to a copper holder of 12.5$\,$mm
width with 2.5$\,$mm channel in its centre. The distance between
the front reflecting surface of the mirror and the MOT's centre
was minimized and equal to 20$\,$mm. The differential tube, formed
by the mirror and its holder, was used both for pumping the volume
of the MOT-source and at the same time for extraction of slow
atoms. An important function of this tube was to reduce the
background flux of thermal atoms. All the cooling laser beams were
produced with five polarization-maintaining fibres and collimation
lenses having a focal length of 150$\,$mm. The resulting diameter
of the beams at $1/e^{2}$ intensity level was about 2.5$\,$cm.
Without active stabilization of the light power after the fibres the intensity was stable to 1$\%.$

The main pair of anti-Helmholtz magnetic coils was oriented along the y-axis. These coils provided
in a centre of the trap the magnetic field gradients of $\nabla B_{y}=8\,$G/cm and $\nabla B_{x}=\nabla B_{z}=-4\,$G/cm
correspondingly. An additional pair of anti-Helmholtz coils compensated
the gradient of the main coils along the z-axis and produced a two-dimensional quadruple field, which was
necessary for realization of the 2D-MOT and the 2D$^{+}$-MOT configurations of the source.

The cooling light, resonant with the $5S_{1/2}(F=2) \rightarrow
5P_{3/2}(F'=3)$ transition of $^{87}$Rb, was provided by a
tapered-amplifier diode laser. The typical power in each of the
four transverse cooling beams was about 12$\,$mW and the power of
the longitudinal cooling beam was about 10$\,$mW. An additional
diode laser, resonant with the $5S_{1/2}(F=1) \rightarrow
5P_{3/2}(F'=2)$ transition, was used for repumping atoms from the
$5S_{1/2}(F=1)$ ground hyperfine state. The power of the
repumping light used in the MOT source was about 5 mW and it was
added to one of the vertical cooling laser beams.

To measure the flux of slow atoms, produced by the MOT-sources, the atoms were trapped by a second 3D-MOT.
The residual gas pressure in the 3D-MOT section of the vacuum chamber was about $10^{-9}\,$mbar.
The measurement of a loading rate of the 3D-MOT directly gave the value of the flux.
The advantage of the method, compared to the approach used in references \cite{Lu,Dieckmann,Conroy,Schoser},
is that it does not rely on how precisely the overall density and the velocity distribution of atoms
in the beam are known. Therefore, the measurement of the flux is as precise as the number of atoms
accumulated in a 3D-MOT is known.
The 3D-MOT was set up at a distance of 34$\,$cm from the source and had a standard six-orthogonal-beams
configuration. The diameters of the laser beams were the same as in the MOT-source and the power of each of them was about
4$\,$mW. A separate repumping laser beam of the same diameter with power of 5$\,$mW was also applied.
The beam of cold atoms entered the 3D-MOT along one of its diagonals. A standard frequency detuning of the 3D-MOT
was the same as for the MOT-source and equal to $\Delta=-3 \Gamma$.
The capture velocity of the 3D-MOT, according to estimates, was above 40$\,$m/s.

Three different configurations of compact MOT-sources were tested.
The LVIS was formed by five laser beams of the correct circular
polarizations. A three-dimensional quadrupole magnetic field of
the LVIS was produced by the main pair of magnetic coils oriented
along the y-axis. The only the difference of the 2D$^+$-MOT was
that the two-dimensional quadrupole magnetic field was used
instead of the three-dimensional one. For that purpose a pair of
anti-Helmholtz coils, oriented along the z-axis (see
Fig.\thinspace 6) and producing a gradient of $\nabla
B_{z}=4\,$G/cm was added. As a result the gradient of the magnetic
field along the z-axis was completely eliminated and the
transverse gradients were equalized to the values of $\nabla
B_{y}=6\,$G/cm and $\nabla B_{x}=-6\,$G/cm. The configuration of
the pure 2D-MOT was the same as the 2D$^+$-MOT, except that the
laser beam along the z-axis was switched off.

The loading curves of the 3D-MOT from the three different
configurations of the MOT-source are presented in Fig.\thinspace
7. Each of the curves there is an average of three subsequent
loading cycles. To calibrate the number of atoms in the 3D-MOT a
standard procedure was applied \cite{Townsend}. For that purpose
an additional circularly polarized probe beam, overlapped with the
centre of the 3D-MOT was introduced. The frequency of the probe
was detuned from the $5S_{5/2}(F=2) \rightarrow 5P_{3/2}(F'=3)$
transition by $\Delta =-3 \Gamma$. Such a detuning of the probe
was chosen to reduce its absorption by the sample of cold atoms.
The residual magnetic field at the centre of the trap in the
absence of the 3D-MOT magnetic field was compensated by optimizing
an optical molasses. The whole sequence of the measurement was as
follows. First, a small number of atoms ($<10^{8}$) was collected
in a 3D-MOT. The loading of the MOT was stopped by switching off
the cooling light of the MOT-source. Then the magnetic field of
the 3D-MOT was switched off and after an additional delay of
10\thinspace ms all cooling beams of the 3D-MOT were also shut
down. At the same moment the probe beam was switched on and the
fluorescence of atoms was measured with a calibrated photodiode.
The repumping light during this measurement stayed on.
The response time of the photodiode was fast enough to detect the
fluorescence signal within first few hundred microseconds, before
atoms were blown away by spontaneous force of the probe light
beam. Based on the formula for the rate of spontaneous scattering
of monochromatic light by a two-level atom and the measured power
of the fluorescence, the total number of atoms in the 3D-MOT was
estimated. In that way the photodiode's signal was normalized to
the number of atoms in the 3D-MOT. This normalization was valid
for a small number of atoms in the MOT, but might not be correct
for the maximum number of loaded atoms, because of absorption of
the MOT beams in the atomic sample. However, the loading atomic
flux is determined only by the loading rate of the MOT at the very
beginning of the process, so the calibration of the photodiode
described above showed the right estimate of the atomic flux. The
Fig.\thinspace 7 shows that the fluxes of LVIS and 2D$^+$-MOT are
nearly equal to one another and are about $8 \times 10^{9}\,$at/s.
This flux provided loading of about $1.3 \times 10^{10}$ atoms
into a 3D-MOT in just three seconds. Although, our calibration, as
it was mentioned above, is not very precise for such large number
of atoms. The pure 2D-MOT provides a much smaller flux of $2
\times 10^{8}\,$at/s. Such a poor flux of a short-length version
of a 2D-MOT source was expected according to the results of our
previous investigations \cite{Schoser}. In absence of cooling
laser beams the source emitted only a background beam of thermal
atoms. The flux of this background thermal beam was so small that
no trapped atoms were detected in the 3D-MOT. This observation
strongly supports the theoretical estimates of this flux to be
two-three orders of magnitude smaller than the flux of cold atoms.

The velocity distribution of atoms extracted from the MOT-source
was measured by a standard time-of-flight (TOF) technique
\cite{Lu,Dieckmann}. A fast switching of the atomic beam intensity
was realized by introducing an additional plug-in laser beam,
which deflected atoms transversely just in front of the
retro-reflecting mirror. This beam had a diameter of 2\thinspace
mm, a power of 10\thinspace mW and was switched with an
acousto-optical modulator. In our experiment, instead of using a
single perpendicular laser beam to detect atoms at some distance
from the source, four beams of the 3D-MOT were used. For that
purpose one pair of couterpropagating beams of the 3D-MOT was
switched off. The other four beams of the MOT, which had an angle
of $\simeq 54^{\circ}$ with respect to the atomic beam, provided a
very fast two-dimensional transverse cooling of atoms and their
concentration at the centre of such a 2D-MOT. To increase the
velocity resolution of the TOF measurement, the diameters of the
laser beams were reduced down to 1\thinspace cm.
The velocity distribution of atoms was extracted from the derivative
of the TOF fluorescence signal. The absolute value of the flux was derived
from the loading of the 3D-MOT as described above.

The resulted velocity distributions of atoms extracted from the LVIS and the
2D$^+$-MOT are presented in Fig.\thinspace 8. One can see that the
velocity of the LVIS (8.5\thinspace m/s) is lower than the
velocity of the 2D$^+$-MOT (11\thinspace m/s). The mean velocities
of atoms extracted from the sources agree remarkably well with the
predictions of our simplified model (see Fig.\thinspace 3). The
FWHM widths of the model velocity distributions were 3.0\thinspace
m/s and 4.5\thinspace m/s correspondingly. The larger widths of
the experimentally measured velocity distributions can be
explained by the stochastic nature of the spontaneous force, which
was not taken into account in the model.

The Fig.\thinspace 9 presents a dependence of the LVIS flux on a total power of cooling laser beams.
At maximum available total laser power of 60\thinspace mW ($\simeq$ 12\thinspace mW per beam) the maximum saturation parameter in each of
the laser beams of the LVIS is $s_{0} \simeq 3$. For the same light intensities the model predicts approximately the same
total flux of atoms of about $8 \times 10^{9}\,$at/s.

In Fig.\thinspace 10 a dependence of the atomic flux on the
imbalance between powers of the two cooling laser beams, directed
along the y-axis is given. One can see that the atomic flux
decreases to less than half if the imbalance factor
$(P_{y-}-P_{y+})/P_{y-}+P_{y+})$ is as small as 0.1. For
simultaneous imbalance in both pairs of transverse laser beams
such a 50$\%$ decrease of the flux is expected at a power
imbalance of 0.05 in each of two pairs of laser
beams, which is in perfect agreement with the prediction of our
model (see Fig.\thinspace 4). The shape of the
experimental dependence  is not as step like as the theoretical prediction of
our model. This can be explained as follows. First, the
model does not take into account the stochastic character of the
light force and collisions between atoms. Second, the experimental
distribution of the light intensity inside the dark extraction
channel might be different from our simplified model. Taking into
account essential absorption losses of laser power in Rb vapors,
at vacuum viewports and retroreflection optics, it is unlikely
that the gain in laser power, achieved by using retroreflection
optics, would compensate for the losses due to intensity
imbalances of the laser beams.

Finally we mention some operational issues related to the running of the different MOT sorces.
There were no essential differences observed between the LVIS and the 2D$^{+}$-MOT sources. Our final choice
was the LVIS, because it does not need the additional compensation coils. The powers of the transverse beams had
to be balanced again, if the output power of the laser changes more than $\pm$ 10$\%$. This can be explained
by the different coupling efficiencies of the fibres.

\section{Conclusion}

It is shown that for well balanced intensities of cooling laser beams the fluxes of slow atoms produced by the LVIS and
2D$^{+}$-MOT sources are nearly the same, but that the velocity of atoms extracted from the 2D$^{+}$-MOT are slightly higher.
The maximum flux $\simeq 8 \times 10^{9}\,$at/s was achieved with laser power of 60\,mW.
Both the experimental measurements and the model presented suggest strong evidence that the observed flux of slow atoms
is the maximum flux achievable for the size and power of cooling
laser beams used.
This flux is 1.5 times higher than the flux of the original LVIS \cite{Lu}, achieved with 500 mW of laser power,
and the same as maximum flux produced by a compact 2D$^{+}$-MOT \cite{Conroy} with laser power of 300\,mW.
One conclusion of this work is that the efficiencies of the LVIS and 2D$^{+}$-MOT sources of slow atoms
in a case of compact configuration are nearly the same and determined mainly by available laser power.
A second conclusion is that the perfect balance of the cooling laser beams of the MOT sources enables one to achieve
the same flux of cold atoms at much lower values of laser power compared to the imbalanced sources.

The compact MOT source will be used in our Rb fountain atomic frequency standard. The achieved flux of
$8 \times 10^{9}\,$at/s allows us to load about $10^{9}$ atoms in just 125\,ms, which is a good starting condition
for an intense Rb fountain.

\section{Acknowledgments}

Many thanks to Dale Henderson, Giuseppe Marra, Witold Chalupczak and Krzysztof Szymaniec for their
technical assistance.
My special thanks to Hugh Klein for his valuable comments to the paper.
This work was funded by the National Measurement System Directorate of the Policy Unit of Trade and Industry, UK.

\newpage

 \begin{figure}
 \includegraphics[scale=0.8]{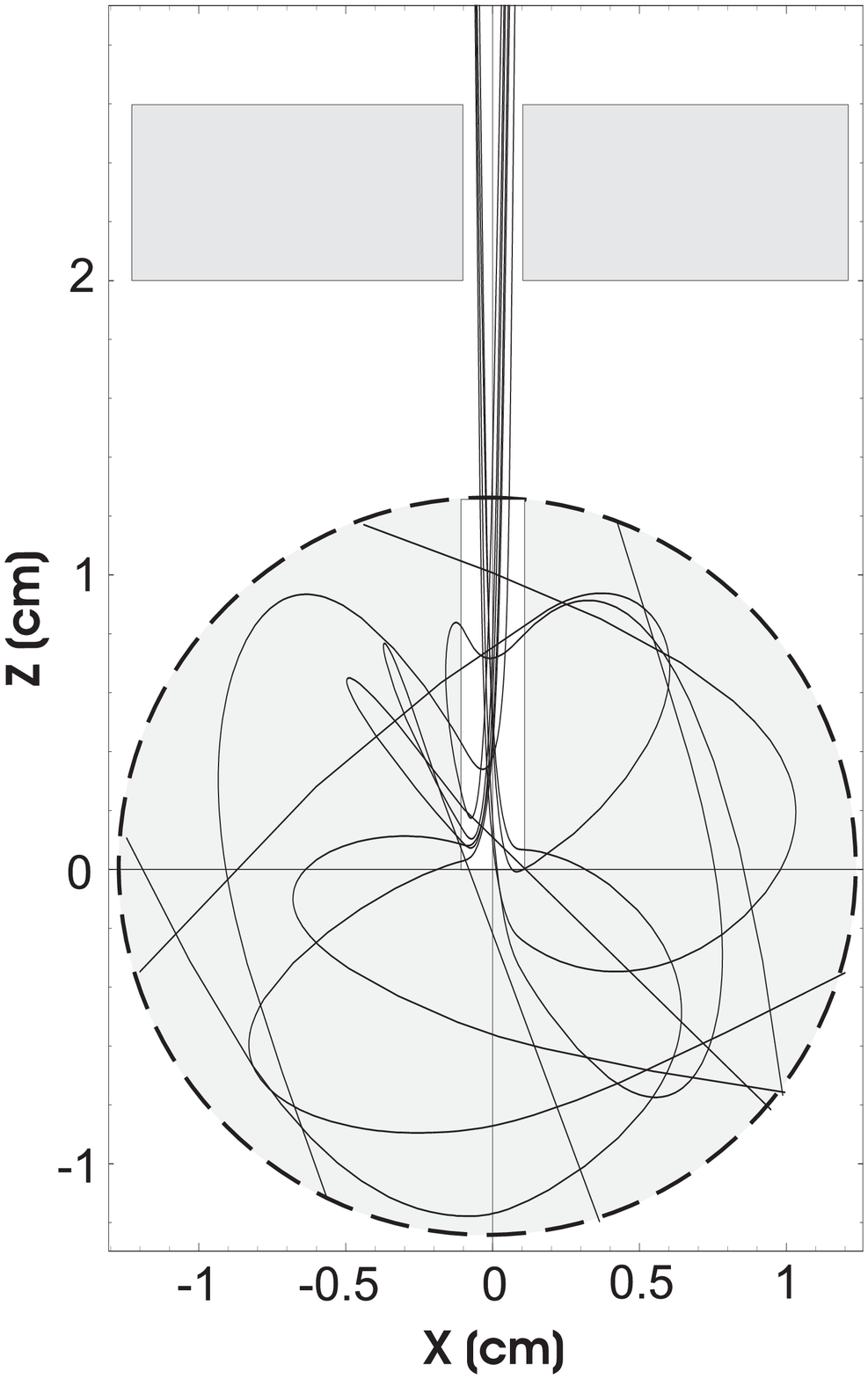}
 \caption{\label{}Trajectories of 10 atoms captured by the LVIS source of slow atoms.
}
 \end{figure}

\begin{figure}
 \includegraphics[scale=0.5]{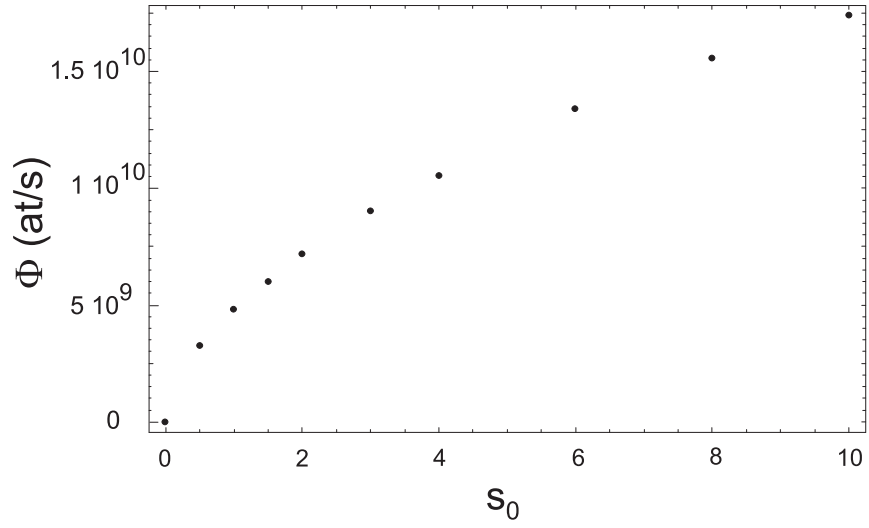}
 \caption{\label{}Calculated dependence of the flux of slow atoms as a function of saturation intensity of the laser
beams of the LVIS source.
}
 \end{figure}

\begin{figure}
 \includegraphics[scale=0.5]{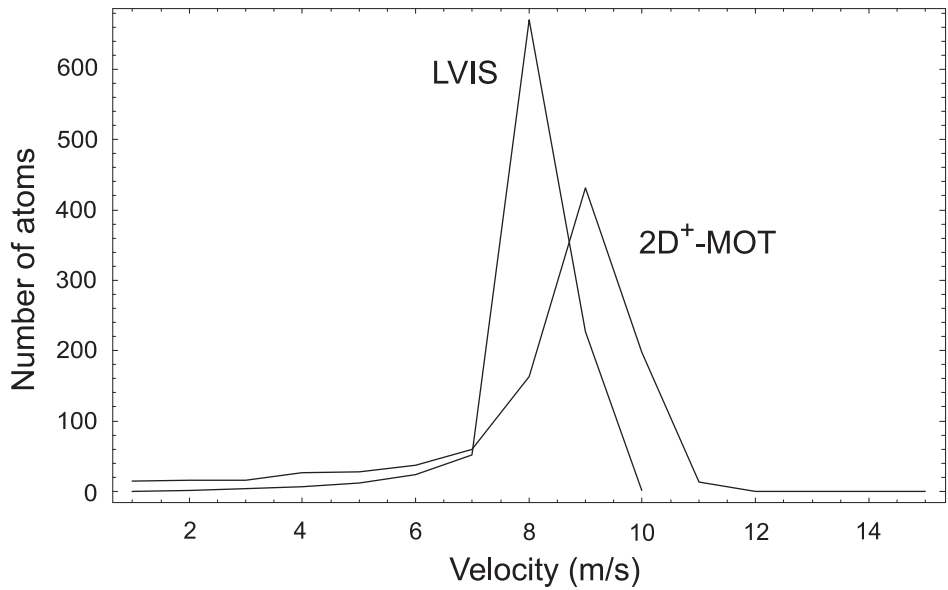}
 \caption{\label{}Calculated velocity distributions of a beam of 1000 slow atoms extracted from the LVIS and the 2D$^+$-MOT.
}
 \end{figure}

\begin{figure}
 \includegraphics[scale=0.5]{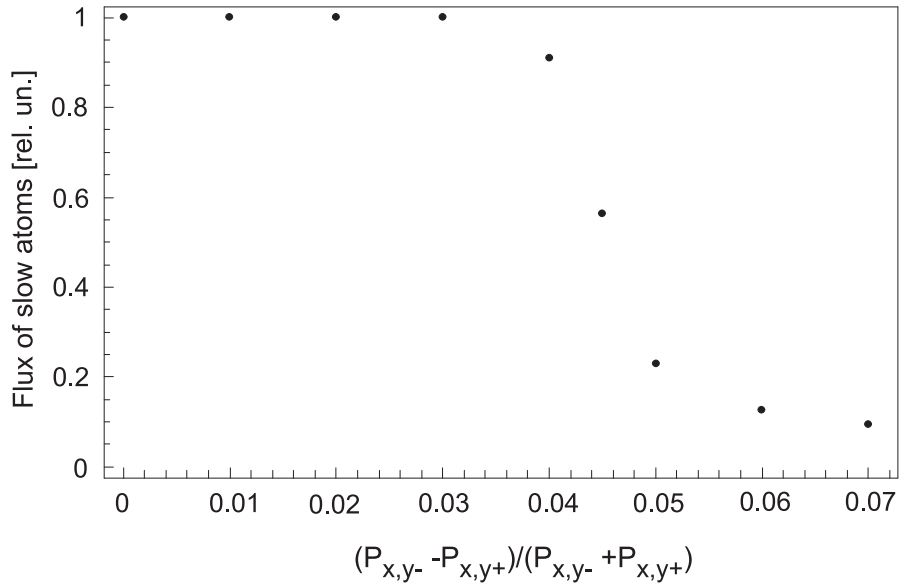}
 \caption{\label{}Dependence of the flux of slow atoms extracted from the LVIS as a function of imbalance between
the two pairs of transverse laser beams.}
 \end{figure}

\begin{figure}
 \includegraphics[scale=0.8]{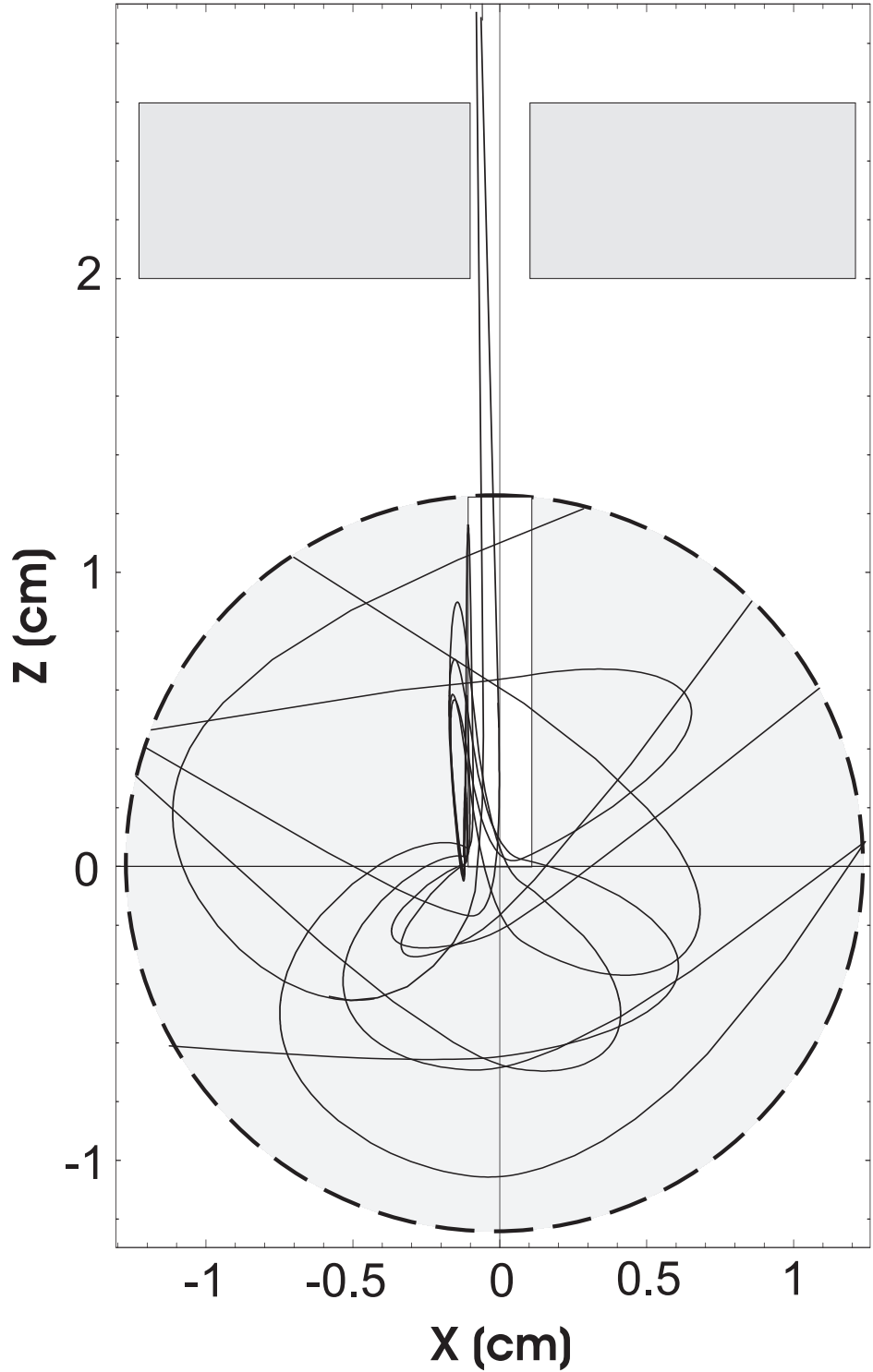}
 \caption{\label{}Trajectories of 10 atoms captured by the LVIS in a case of 5\% power imbalance of the
transverse cooling laser beams.}
 \end{figure}

\begin{figure}
 \includegraphics{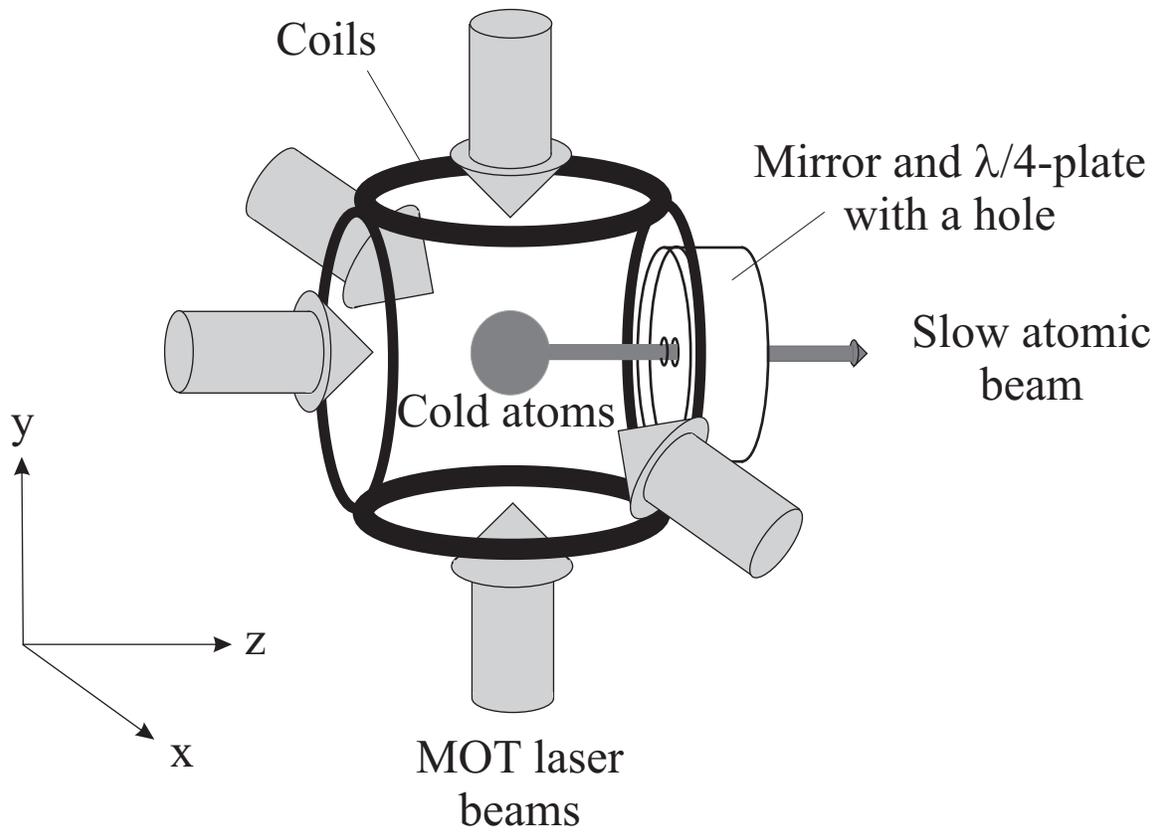}
 \caption{\label{}Schematic of a MOT-source of slow atoms.
}
 \end{figure}

\begin{figure}
 \includegraphics[scale=0.5]{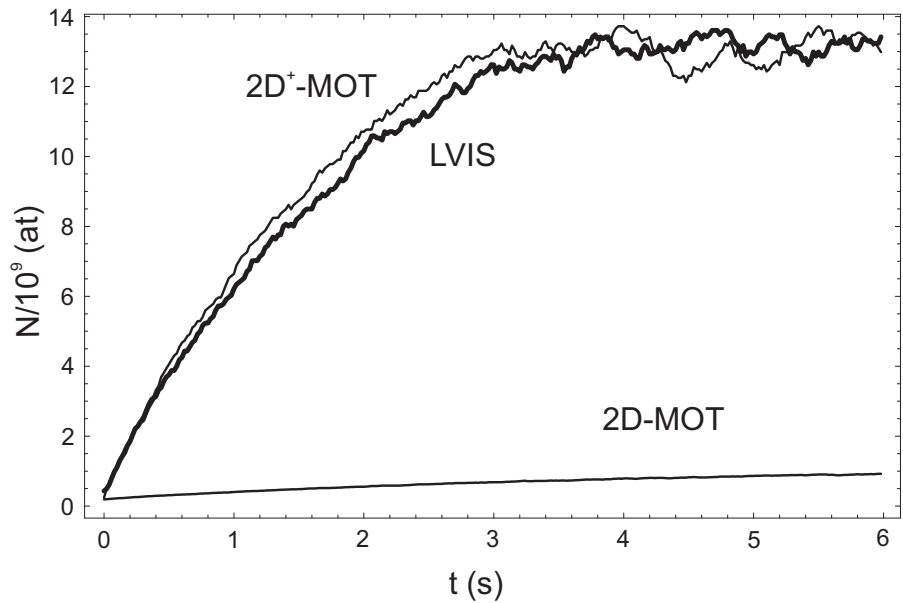}
 \caption{\label{}Loading of atoms into the 3D-MOT from the LVIS, 2D$^+$-MOT and pure 2D-MOT sources as a function of time.
}
 \end{figure}

\begin{figure}
 \includegraphics[scale=0.5]{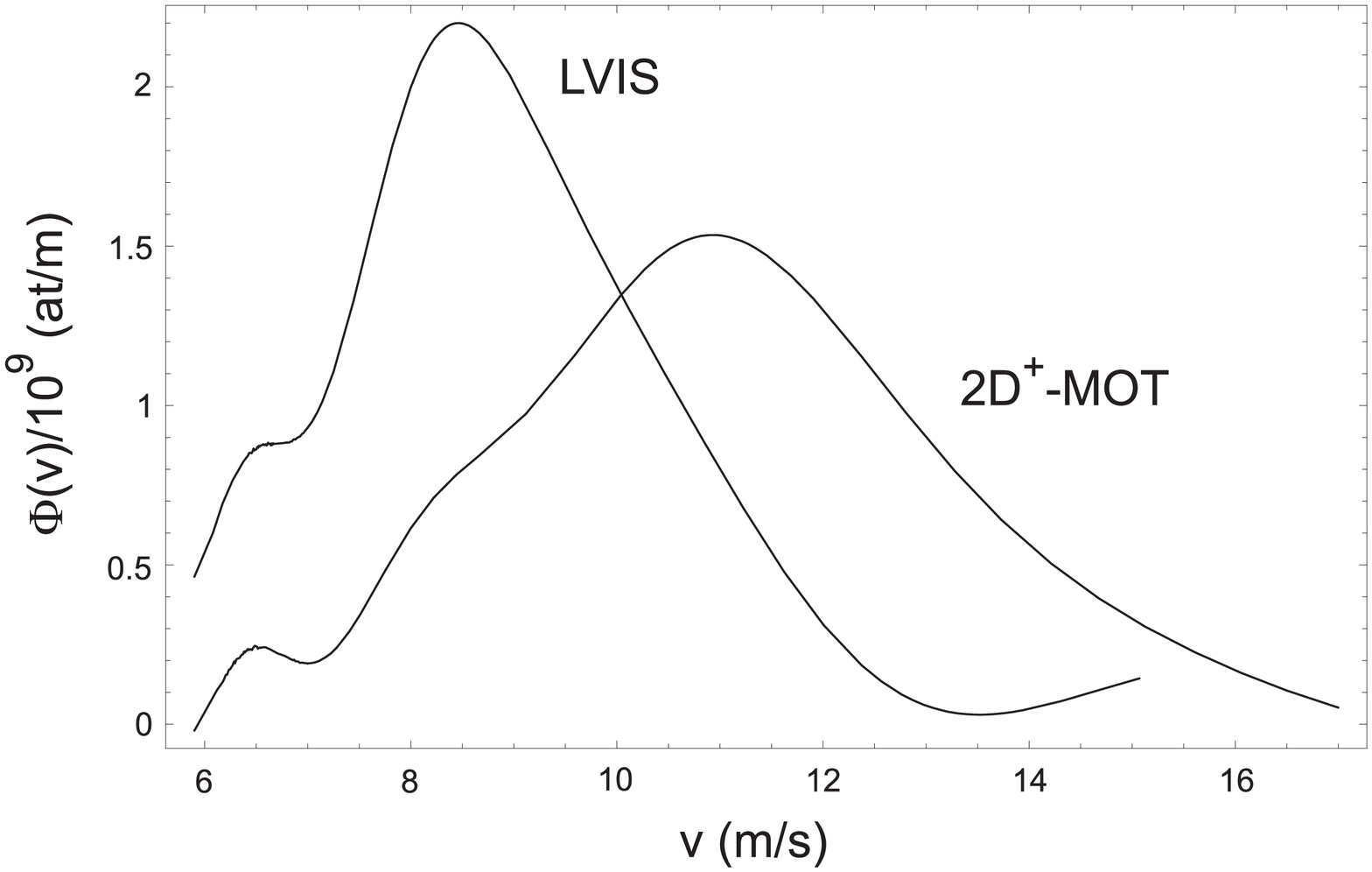}
 \caption{\label{}Experimental velocity distribution of the beam of slow atoms extracted from the LVIS and 2D$^+$-MOT sources.
}
 \end{figure}

\begin{figure}
 \includegraphics[scale=0.5]{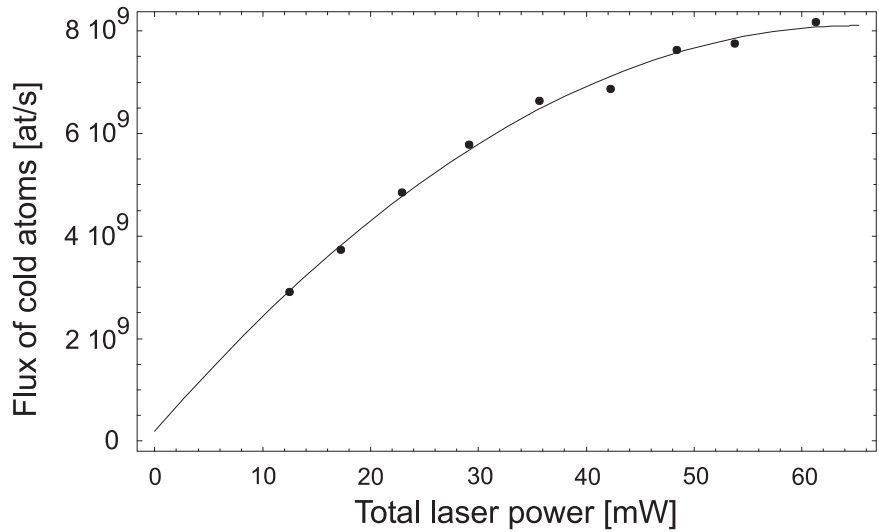}
 \caption{\label{}Dependence of the total flux of slow atoms of the LVIS as a function of the total laser power.
}
 \end{figure}

\begin{figure}
 \includegraphics[scale=0.5]{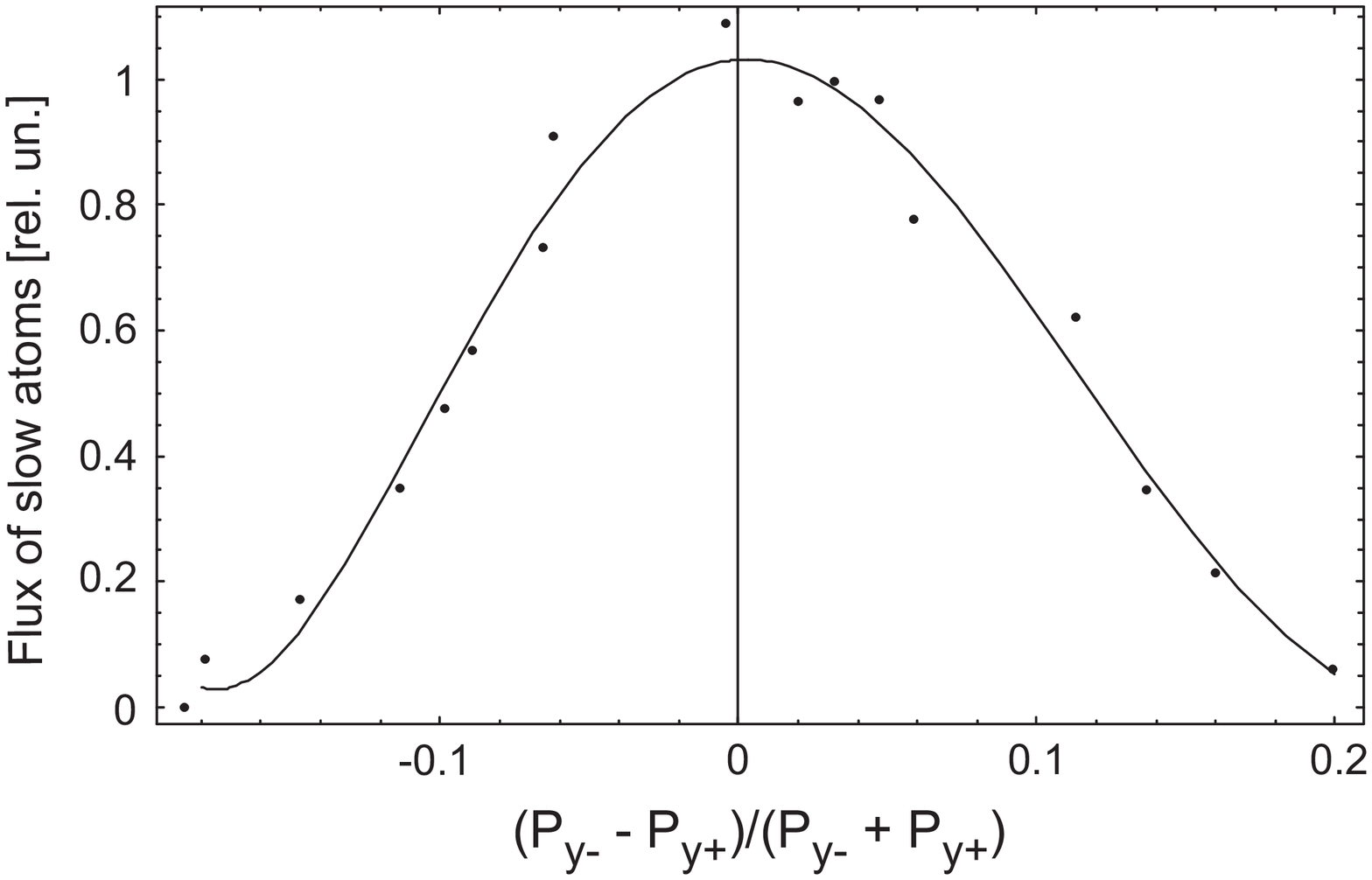}
 \caption{\label{}Dependence of the flux of slow atoms extracted from the LVIS as a function of imbalance between
the pair of vertical transverse laser beams.
}
 \end{figure}


\bigskip

\begin{references}
\bibitem{Raab}  E. Raab, M. Prentiss, A. Cable, S. Chu, and D. Pritchard, Phys. Rev. Lett. {\bf 59} (1987)
2631.

\bibitem{Roos}  C.F. Roos, P. Cren, J. Dalibard, D. Cury-Odelin, Physica Scripta {\bf T105} (2003) 19.

\bibitem{Conroy}  R.S. Conroy, Y. Xiao, M. Vengalattore, W. Rooijakkers, M. Prentiss, Opt. Comm.
{\bf 226} (2003) 259.

\bibitem{Phillips}  W. Phillips, H. Metcalf, Phys. Rev. Lett. {\bf 48} (1982) 596.

\bibitem{Cornell}  E.A. Cornell, C.E. Wieman, Intern. Journ. of Mod. Phys. B {\bf 16} (2002) 4503.

\bibitem{Lu}  Z.T. Lu, K.L. Corwin, M.J. Renn, M.H. Anderson, E.A. Cornell, and C.E. Wieman,
Phys. Rev. Lett. {\bf 77} (1996) 3331.


\bibitem{Dieckmann}  K. Dieckmann, R.J.C. Spreeuw, M. Weidem\"{u}ller, and J.T.M. Walraven, Phys. Rev. A {\bf 58}
(1998) 3891.

\bibitem{Schoser}  J. Sch\"{o}ser, A. Bat\"{a}r, R. L\"{o}w, V. Schweikhard, A. Grabowski, Yu.B. Ovchinnikov,
and T. Pfau, Phys. Rev. A {\bf 66} (2002) 023410.

\bibitem{Nesmeyanov} A.N. Nesmeyanov, Vapor Pressure of the Chemical Elements, Ed. by Robert Gary, Elsevier, Amsterdam, 1963.

\bibitem{Townsend} C.G. Townsend, N.H. Edwards, C.J. Cooper, K.P. Zetie, C.J. Foot, A.M. Steane, P. Szriftgizer, H. Perrin,
J. Dalibard, Phys. Rev. A {\bf 52} (1995) 1423.

\end{references}
\end{document}